\documentclass[a4paper,twoside,english]{paper}
\usepackage[T1]{fontenc}
\usepackage[latin9]{inputenc}
\usepackage{float}
\usepackage{url}

\makeatletter

\floatstyle{ruled}
\newfloat{algorithm}{tbp}{loa}
\floatname{algorithm}{Algorithm}

\makeatother

\usepackage{babel}

\begin{document}

\title{A Monte Carlo Algorithm for Universally Optimal Bayesian Sequence
Prediction and Planning}

\author{Anthony Di Franco}

\date{01/03/2010}

\maketitle

\part{Algorithm}

\section*{Abstract}

The aim of this work is to address the question of whether we can
in principle design rational decision-making agents or artificial
intelligences embedded in computable physics such that their decisions
are optimal in reasonable mathematical senses. Recent developments
in rare event probability estimation, recursive bayesian inference,
neural networks, and probabilistic planning are sufficient to explicitly
approximate reinforcement learners of the AIXI style with non-trivial
model classes (here, the class of resource-bounded Turing machines).
Consideration of the effects of resource limitations in a concrete
implementation leads to insights about possible architectures for
learning systems using optimal decision makers as components.

\section*{Keywords}

recurrent neural network, Turing machine, AIXI, reinforcement learning,
planning, sequence prediction, Universal prior

\section{Introduction}

Hutter \cite{Hutter:04uaibook} defines the universal algorithmic
agent AIXI, a Pareto-optimal policy for arbitrary computable stochastic
environments. Here we consider a Monte Carlo algorithm for sequence
prediction and planning in a Turing-universal model class which seeks
to statistically approximate resource-bounded AIXI while avoiding
the enumerative search in similarly powerful algorithms replacing
it with statistical approximations to the governing probability distributions.

Hutter held that {}``AIXI is computationally intractible, however,
AIXI can serve as a gold standard for A{[}rtificial{]} G{[}eneral{]}
I{[}ntelligence{]} and AGI research.'' \cite{Hutter:09agitalk} The
main difficulty in computing AIXI in a universal model class is obtaining
the algorithmic probabilities of the hypotheses, since the algorithmic
probabilities are only limit-computable. Previous approaches to learning
in this class enumerate proofs or enumerate and run programs in a
hypothesis space with their executions interleaved \cite{Schmidhuber:04oops},
\cite{Schmidhuber:09gm}. We take a different approach characterized
by two main points: 1) The computation runs in tandem with the environmental
process, and to succeed must produce the correct result at the same
time the environment demands it. Thus there is an implicit {}``speed
prior'' \cite{Schmidhuber:02speed} imposed on the model parameterized
by the frequency of operation of the model with respect to the environment,
or, alternatively, we are approximating AIXI-$tl$ with a time bound
equal to the relative frequencies of the model's cycle and the environmental
inputs, and length bound determined by the space available for the
model representation, but we rely directly on statistically valid
approximations rather than proofs of valid approximation. 2) Otherwise,
hypotheses are weighted according to their representation-length complexity
as in AIXI.

This part presents the algorithm, with experimental results deferred
to the second part, which is in preparation.

\section{Architecture}

Using the equivalence between recurrent neural networks and Turing
machines \cite{Hyotyniemi:96tm}, we choose recurrent neural networks
as the model class. Additional information on the relationships between
recurrent neural network architectures and computational classes is
available in \cite{Seigelmann:99nn}. The main advantages of this
representation are that 1) it is parameterized by vectors of dense
numbers and therefore straightforward to sample from in a meaningful
and statistically efficient way using established Monte Carlo techniques,
and 2) while recurrent neural networks embed the class of Turing machines,
they can be counted on to produce meaningful outputs at any point
in their operation, so do not suffer from the halting problem in a
way that presents algorithmic difficulties.

Among the class of recurrent neural networks, we choose a modified
version of Long Short-Term Memory \cite{LSTM} because it has established
state-of-the-art performance on benchmark tasks and because it can
easily be seen as a continuously-parameterized flip-flop fed and controlled
by perceptrons, making the the interpretation of the model more straightforward,
and constraining it to perform in a way known to be useful. This is
a different motivation from its original one which was to prevent
the premature decay of error signals in training. The modification
here is to add a {}``bypass'' gate which mixes the computation directly
into the output, rather than forcing the computation to first be stored
in the state before being output, so that a cell can act either as
a memory or as a computational element according to circumstance.

The network outputs are made to predict the network inputs at the
next time step. Network inputs are presented in a binary encoding,
so that network outputs can be treated directly as a probability (that
of observing a set input bit) without less general assumptions on
the probability distributions of network inputs.

To plan, we present as input to the network a concatenation of the
observed environmental inputs (of which some subset is interpreted
as a reward signal) and the agent's own outputs, and make the network
generate samples of likely continuation sequences by predicting a
distribution over inputs and outputs for the next timestep, sampling
from the predicted distribution, predicting again by taking the sample
as if it had been observed, and so on recursively, and generated samples
from likely continuation sequences to plan by generating them out
to a specified horizon and selecting the next action in the continuation
that yields the most aggregate reward in the sample.

\section{Algorithm}

\subsection*{Overview}

We first obtain a sample from the Universal prior over the model class.
This must be done only one for a certain choice of model class (i.e.
network topology). Thereafter, the algorithm operates in discrete
timesteps. At each timestep, the current environmental inputs and
the agent's most recent outputs are presented as inputs to the network,
and the network updates its estimate of the neural network parameters
according to the probabilities it previously assigned to the observed
inputs. If an action is called for at the current timestep, the current
estimate of the network parameters is used to generate likely continuation
sequences out to a specified time horizon, generating a likely continuation
of the sequence of network inputs, which consist of (environmental
input, agent output) pairs and where the environmental input contains
a reward signal. The next action is chosen as the agent output that
serves as a partial prefix to the set of continuation sequences with
the most expected reward.

\subsection{Prediction}

Prediction has two components, the first being obtaining a useful
sample from the Universal prior, and the second being recursively
updating that sample to reflect the posterior distribution in light
of the data. For the sample from the Universal prior, we face the
practical difficulty that we wish to have significant numbers of samples
from regions of the sample space with exponentially decreasing probability,
and so turn to techniques from rare-event probability estimation.
For the computation of the posterior probability, we rely on standard
techniques from recursive Bayesian inference.

We seek to sample models from the model class with large weights $w$
under the following distribution so as to use $\xi$, the Universal
posterior, as our predictor, that is, we take the prior probabilities
of hypotheses to be their Universal probabilities and maintain their
posterior likelihoods by Bayes' rule (equations and notation from
\cite{Hutter:04uaibook}):

\[
w_{\nu}(x_{t=0})=2^{-K(\nu)}\]

\[
\xi(x_{t}|x_{<t})=\sum_{\nu\in\mathcal{M}}w_{\nu}(x_{<t})\nu(x_{t}|x_{<t}),w_{\nu}(x_{1:t})\leftarrow w_{\nu}(x_{<t})\frac{\nu(x_{t}|x_{<t})}{\xi(x_{t}|x_{<t})}\]

We measure the encoding-length (Kolmogorov) complexity of the mapping
from current states and inputs to subsequent states and outputs with
the $B$ quantity from Flat Minimum Search \cite{FMS} which gives
the number of bits required to encode the network's parameters to
maintain a certain amount of precision in the mapping it represents.
In the general case of multiple inputs going to multiple outputs,
the full $B$ quantity is required, but in this implementation where
each weight directly influences only one output within an iteration,
the following simplification obtains, (with a constant term omitted
as being irrelevant,) a quantity reminiscent of the Fisher information:

\[
B=\sum_{w\in weights}\log(\frac{\partial output_{w}}{\partial w})^{2}\]

We might also do this by following up to the determination of the
gradient the RTRL algorithm, or an equivalent technique given a different
choice of RNN architecture, in which case we might then maintain the
prior and posterior terms in the $\xi$ weighting separately, and
recompute the prior term with the recursive formula for the RTRL partial
derivatives, and update the posterior term recursively according to
the formula given previously for $\xi$. However, the previous paragraph's
method is most appropriate to the interpretation of the recurrent
neural network as a Turing machine, with the states serving as a bounded
work tape, and the neural network map serving as the state-transition
dynamics, and is preferred here.

Using $B(\nu)$ to approximate $K(\nu)$, we initialize a sample set
of size $S$ by rare-event probability estimation by strata and acceptance
/ rejection according to the Universal prior distribution parameterized
by $B$. That is, given a proportion $P$ between zero and one, say
one half, we generate samples from $\nu\in\mathcal{M}$ according
to $p(\nu)=2^{-B(\nu)+K(x_{\nu})}$ with Metropolis-Hastings where
$K(x_{\nu})$ is the encoding complexity of a state vector sampled
jointly with the weights of $\nu$, then within the upper $P$th quantile
of the sample, we construct a kernel density estimator and recursively
repeat the $P$th quantile sample with the quantile, recording the
boundary value of the quantile at each step of the recursion as a
sequence of levels. Given the sequence of levels, we repeat the process
with the level thresholds replacing the quantiles. Notably, we obtain
from this process an estimate of the normalization constant of the
Universal distribution, though it is unused here. See \cite{Botev:08count}.

\begin{algorithm}
\caption{Sampling from the Universal Prior}

1. Choose a sample size $S$, number of levels desired $N_{L}$, and
quantile $P$; suggested are $S=2^{parameterSpaceCardinality}$, $N_{L}=parameterSpaceCardinality$,
$P=2$.

2. Generate an initial set of samples with Metropolis-Hastings by
repeatedly drawing from a (say) multivariate Gaussian or Laplacian
proposal centered at the current sample, accepting samples as their
bounded likelihood ratio $\min(\frac{2^{-B(\nu')+K(x_{\nu'})}}{2^{-B(\nu)+K(x_{\nu})}},1)$
is greater than a draw from a uniform(0,1) random variable.

3. Proceed with algorithms 2.2 and 2.1 of \cite{Botev:08count} with
a transition kernel $g*$ given by a kernel density estimator over
the sample, suggested is the multivariate Gaussian or Laplace distribution
with the sample variances or covariances.
\end{algorithm}

Having thus obtained a sample from the Universal prior, which must
only be done once per instantiation of a network of a certain architecture,
we then update the posterior term recursively according to the formula
given previously for $\xi$. 

Upon a degeneration of the sample as indicated by the variance of
the weights falling below a given threshold, we may resample by replicating
and replacing samples with samples from a kernel density estimate
according to their weights as in the sequential Monte Carlo literature,
especially \cite{Balakrishnan06aone-pass}. We can include in each$\nu$
a representation of the state vector, that is, sample states and dynamics
as a block, or maintain separate estimates conditional on one another
as is more common in Kalman-filtered recurrent neural networks. Here
we choose the latter. We may choose to use multiple RNN simulation
steps per time step of the environment to relax the time bound on
the computation. Our estimate of the probability of an input bit being
set is then given by the $w_{\nu}$-weighted sum of our samples from
$\mathcal{M}$.

\begin{algorithm}

\caption{Recursive Bayesian Posterior Update}

Denote the current observation by $x_{t}$, the previous set of weights
by $w_{\nu}(x_{<t})$, then:

1. Compute the probability of each sample $\nu$ in the sample set
by propagating each of the current weighted state samples through
the network evolution represented by $\nu$, and taking the likelihood
of the resulting multivariate Bernoulli with respect to the observed
bits;

2. Update the weights using the resulting likelihoods according to
$w_{\nu}(x_{1:t})\leftarrow w_{\nu}(x_{<t})\frac{\nu(x_{t}|x_{<t})}{\xi(x_{t}|x_{<t})}$.

3. If the effective sample size $\frac{(\sum_{\nu}w_{\nu})^{2}}{\sum_{\nu}(w_{\nu})^{2}}$
has fallen below a pre-determined threshold, say 50\%, then:

3a. Sample with replacement from the set of $\nu$ with probabilies
$w_{\nu}$ and assign each a unit weight to form a new sample with
the same distribution;

3b. Form a kernel density estimator for the new sample and take a
new sample from it, assigning each unit weight. Suggested is the shrunk
kernel density estimator given, for a kernel bandwidth $b$, by computing
$a=\sqrt{1-b^{2}}$, translating each parameter $\Theta_{\nu}$ in
each sample $\nu$ towards the sample mean $\bar{\Theta}$ by taking
the linear combination $a\Theta_{i}+(1-a)\bar{\Theta}$, then sampling
from a kernel parameterized by the translated particles and a covariance
of $b^{2}V$, $V$ being the sample variance matrix. (See \cite{Balakrishnan06aone-pass})

4. Sample a new set of states by repeating steps 2 and 3 with the
roles of the network parameters and states reversed.
\end{algorithm}

\subsection{Planning}

This model is observable, and state-determined, so the prerequisites
for the application of the approximately optimal stochastic planning
algorithm in \cite{kearnsmansourng} are satisfied. To plan, we jointly
model environmental inputs, including the reward signal, and agent
actions, and sample continuations of the joint sequence of these.
We obtain a distribution over the future rewards of certain actions
or action sequences by grouping continuation samples according to
the actions or action sequences they contain, and plan by selecting
at each time step the next action with the highest expected total
future reward. Arbitrary discounting is easily accomodated by applying
the appropriate factor to the sum at each step in the continuation
sequence. As an optimization, if the environment is sufficiently stationary
and sufficiently well-approximated by the model, it is conceivable
that prefixes longer than one timestep in length could be chosen and
executed without re-running the planner, but this is not considered
further here.

\begin{algorithm}

\caption{Planning}

At each time step when an action is called for:

1. Generate a probability distribution over continuations of network
input sequences by making a copy of the state samples, and iterating
steps 2 - 4 of algorithm 2 on the state copy but without modifying
the estimate of the network parameters, until a sample of continuation
sequences has been produced out to the desired time horizon.

2. Group the continuation sequences according to their sharing a common
next agent action, and take the expected discounted future reward
within each group.

3. Perform the action with the highest expected discounted future
reward.
\end{algorithm}

\subsection{Remarks}

The algorithmic-complexity-based parameter smoothing accomplishes
a form of principled dimensionality reduction within the upper bound
imposed by the dimension of the representation. A reversible-jump
Monte Carlo sampler could be used to select the dimensionality without
an explicit upper bound \cite{deFreitas:rj}.

To model input-output relationships, input-output pairs should be
presented as network inputs, with use of the planner being unnecessary,
although the more typical approach analogous to the operation of other
neural networks of giving the input as the network input, calling
for the output as the agent action, and giving a reward signal inversely
related to the output error should also work, but waste planning effort.

\subsection{Implementation notes}

A prototype implementation is available at \url{http://code.google.com/p/machine-tools}
in the SolomonoffNet module.

We compute the partial derivatives of network outputs with respect
to weights, the key input to the computation of $B$, using forward-mode
automatic differentiation, a choice made mostly for implementation
convenience and because in the general case here, where input dimension
equals (or is less than) output dimension, simply-implemented forward-mode
differentiation is as efficient as (resp. more efficient than) the
reverse-mode (backpropagation) more commonly encountered in neural
network algorithms.

\section{Outlook}

This paper applies sampling to effect AIXI-style learning in a specific
Turing-universal model class, but analogous techniques apply to any
model class where the parameters can be sampled, with appropriate
modifications to take into account the needs of sampling from another
such model class, particularly the determination of the encoding complexity,
and the resulting mixture estimate is as close to the true distribution
as the model class permits \cite{Hutter:04uaibook}.

In the recurrent network setting the map encoding complexity of the
system dynamics $B$ is equivalent to the information rate of the
corresponding process and the topological entropy of the corresponding
dynamical system. Together with the sampling method presented here
for the Universal prior, it could be used to numerically characterize
the complexity of other classes of dynamical systems.

Approximately, but neglecting the encoding complexity structure of
the parameter space, the size of the parameter space to be sampled
grows exponentially with network dimension, as does thus the cost
associated with maintaining the sample set, and the costs of other
implementation details such as the Jacobian accumulation and the peephole
connections across modules in LSTM grow faster than linearly. To mitigate
these costs, a large network may be constructed as a set of small
network modules with the ability to view one another\textquoteright{}s
outputs or states as additional environmental inputs, and a perhaps
probabilistic locality structure can be imposed to limit the interconnection
costs while maintaining bandwidth among computational nodes in the
network, such as a topology with power-law node degrees. An analogy
can be drawn to decoupled extended Kalman filter training of recurrent
neural networks, though the interactions of the modules require a
more powerful theoretical framework to analyze because of the bounded-Universal
power of the models. Computability Logic \cite{japaridze:logic} may
be such a framework, wherein the network modules can take each other
as oracles. Additionally, one can consider allowing these modules,
capable of arbitrary computation, to interact with other special-purpose
modules, such as smoothers of system input / output, pattern-associative
memories, chess programs, etc. depending on the problem domain.

This motivation for decoupling the large network into modules and
to view interactions of modules with their environment in terms of
computabilty logic suggests a model for the mammalian brain where
cortical columns are identified with resource-bounded general-purpose-computation
modules and other parts of the brain are either computational oracles
or filters on input / output with the environment. Suppose that the
thalamus serves as the communication network by which the modules
are connected with one another over long ranges and with other oracles
/ filters in the brain and the (perhaps filtered) environmental inputs
/ agent controls, and is responsible for distributing reward signal
information across the computational modules; that the hippocampus
serves as a pattern-associative memory in tandem with the cortical
ensemble; that the cerebellum serves as motor output consolidation
and smoothing. Working under that hypothesis, the functional specialization
of different regions of cortex might be derived from considerations
of the computational capacity of individual modules and the information
bandwidth of the connections between groups of modules in a region,
other regions relevant to its function, and the locations in the network
of the relevant environmental inputs / outputs, rather than any sort
of a priori enforced anatomical specialization. Also, the brain can
thus be seen to implement a solution to the prediction and planning
problems that is optimal in a resource-bounded, stochastic sense,
and quite general.

Separately, but also based on the idea of an architecture for an intelligent
agent built around a communication network that transmits data from
the environment to a set of computational units, permits these computational
units to communicate with one another, and transmits commands back
to be executed upon the environment by the agent's effectors, this
agent architecture suggest a means of quantifying Boyd's Observe-Orient-Decide-Act
loop conceptual framework \cite{boyd} for analyzing strategic interactions,
where observation is the capacity in units of bits per second to provide
sensory information to the computational units, orientation is the
capacity of the computational units to inform one another of computed
characterizations of the environment and is bounded by their speed
of computation and the capacity of the communication network among
them measured in bits per second, decision is equated with the agent's
planning capacity as manifested by the sequence of outputs it produces
and the capacity of the communication network to communicate them
to the effectors and is also measured in bits per second, and action
is equated with actuation's ability to affect the environment in units
of joules per second. Thus Boyd's Energy-Maneuverability theory can
also be quantified in units of bit-Joules per second. It also suggests
a potential approach to making mathematically rigorous his references
to Heisenberg's uncertainty principle, Goedel's incompleteness theorem,
and the second law of thermodynamics, which were merely illustrative
in his presentation of his ideas, by formulating analogous statements
within the context of statistics and algorithmic probability of AIXI.

Schmidhuber's history-compressing neural networks \cite{schmidhuber:histcomp}
implemented, in a sense, a less general form of computational modules
using one another as oracles, where the modules were arranged in a
heirarchy and learned abbreviated representations of ever-longer regular
input sequences.

\section*{Notes}

This work presents a refined version of the algorithm described in
the extended abstract {}``Overview of A Monte Carlo Algorithm for
Universally Optimal Bayesian Sequence Prediction and Planning'' \cite{dfko:unpub}.

\bibliographystyle{mdpi}
\bibliography{net}

\end{document}